\documentclass[reprint, double column, superscriptaddress,prl, showkeys]{revtex4-1}

\usepackage{graphicx,epsfig}
\usepackage{amssymb}
\usepackage{amsmath}
\usepackage{bm}
\usepackage{textcomp}
\usepackage{color}
\usepackage{pgffor}

\usepackage{soul}

\usepackage{pdfpages}
\usepackage{BOONDOX-cal}

\makeatletter

\AtBeginDocument{\let\LS@rot\@undefined}
\makeatother

\newif\ifarXiv
\arXivtrue

\begin{document}
\setcounter{page}{1}

\title[]{Even-Denominator Fractional Quantum Hall State at Filling Factor $\nu=3/4$}
\author{Chengyu \surname{Wang}}
\author{A. \surname{Gupta}}
\author{S. K. \surname{Singh}}
\author{Y. J. \surname{Chung}}
\author{L. N. \surname{Pfeiffer}}
\author{K. W. \surname{West}}
\author{K. W. \surname{Baldwin}}
\affiliation{Department of Electrical and Computer Engineering, Princeton University, Princeton, New Jersey 08544, USA}
\author{R. \surname{Winkler}}
\affiliation{Department of Physics, Northern Illinois University, DeKalb, Illinois 60115, USA}
\author{M. \surname{Shayegan}}
\affiliation{Department of Electrical and Computer Engineering, Princeton University, Princeton, New Jersey 08544, USA}
\date{\today}

\begin{abstract}

Fractional quantum Hall states (FQHSs) exemplify exotic phases of low-disorder two-dimensional (2D) electron systems when electron-electron interaction dominates over the thermal and kinetic energies. Particularly intriguing among the FQHSs are those observed at even-denominator Landau level filling factors, as their quasi-particles are generally believed to obey non-Abelian statistics and be of potential use in topological quantum computing. Such states, however, are very rare and fragile, and are typically observed in the excited Landau level of 2D electron systems with the lowest amount of disorder. Here we report the observation of a new and unexpected even-denominator FQHS at filling factor $\nu=3/4$ in a GaAs 2D \textit{hole} system with an exceptionally high quality (mobility).  Our magneto-transport measurements reveal a strong minimum in the longitudinal resistance at $\nu=3/4$, accompanied by a developing Hall plateau centered at $(h/e^2)/(3/4)$. This even-denominator FQHS is very unusual as it is observed in the \textit{lowest} Landau level and in a 2D \textit{hole} system. While its origin is unclear, it is likely a non-Abelian state, emerging from the residual interaction between composite fermions.

\end{abstract}

\maketitle  

Since its discovery in 1982 \cite{Tsui.PRL.1982}, the fractional quantum Hall effect has been one of the most active topics in condensed matter physics \cite{Halperin.Jain.Book.2020}. It is observed in low-disorder two-dimensional electron systems (2DESs) at low temperatures and large, quantizing, perpendicular magnetic fields, when electrons' thermal and kinetic energies are quenched and the Coulomb interaction between the electrons dominates. The vast majority of fractional quantum Hall states (FQHSs) are observed in the lowest Landau level (LL) at \textit{odd-denominator} LL filling factors, and can be mostly understood in a standard composite fermion (CF) model \cite{Halperin.Jain.Book.2020, Jain.PRL.1989, Halperin.PRB.1993, Jain.book.2007}. By attaching $2m$ flux quanta to each electron, the FQHSs at $\nu=p/(2mp\pm 1)$, the so-called Jain-sequence states, can be mapped to the integer quantum Hall states at the Lambda level ($\Lambda$L) filling factor $p$ of the weakly interacting $2m$-flux CFs ($^{2m}$CFs). 

Thanks to intense experimental efforts over the last few decades and improvements in sample quality (mobility), new FQHSs which cannot be explained in the standard Jain sequence have been reported \cite{Halperin.Jain.Book.2020, Willett.PRL.1987, Goldman.Shayegan.Surf.Sci.1990, Suen.PRL.1992, Eisenstein.PRL.1992, Suen.PRL.1994, Pan.PRL.1999, Pan.PRL.2003, Xia.PRL.2004, Luhman.PRL.2008, Pan.PRB.2008, Shabani.PRL.2009, Shabani.PRL.2009.asym, Bellani.PRB.2010, Shabani.PRB.2013, Liu.PRL.2014.1/2, Liu.PRB.2014, Pan.PRB.2015, Samkharadze.PRB.2015, Mueed.PRL.2015, Mueed.PRL.2016}. Among these are FQHSs observed at certain \textit{even-denominator} fillings, e.g. at $\nu=5/2$ \cite{Willett.PRL.1987}.  Although its origin is not yet entirely clear, theory \cite{Haldane.PRL.1988, Moore.NPB.1991, Morf.PRL.1998} strongly suggests that the $\nu=5/2$ FQHS is a spin-polarized, $p$-wave paired (Pfaffian) state with non-Abelian statistics, rendering it a prime candidate for fault-tolerant, topological quantum computing \cite{Nayak.RMP.2009}. Even-denominator FQHSs have also been reported at other filling factors, e.g., at $\nu=1/2$ and $1/4$ in wide GaAs quantum wells \cite{Suen.PRL.1992, Suen.PRL.1994, Luhman.PRL.2008, Shabani.PRL.2009, Shabani.PRL.2009.asym, Shabani.PRB.2013, Liu.PRL.2014.1/2, Liu.PRB.2014, Mueed.PRL.2015, Mueed.PRL.2016}. The origin of these states is also unclear: some experimental and theoretical results are consistent with these being two-component, Halperin-Laughlin, Abelian states \cite{Suen.PRL.1994, Shabani.PRL.2009, Shabani.PRB.2013, Peterson.PRB.2010}, although the latest data and calculations suggest a one-component, Pfaffian, non-Abelian origin \cite{Shabani.PRL.2009.asym, Mueed.PRL.2016, Zhu.PRB.2016, Faugno.PRL.2019, Zhao.PRB.2021}.

\begin{figure*}
  \begin{center}
    \psfig{file=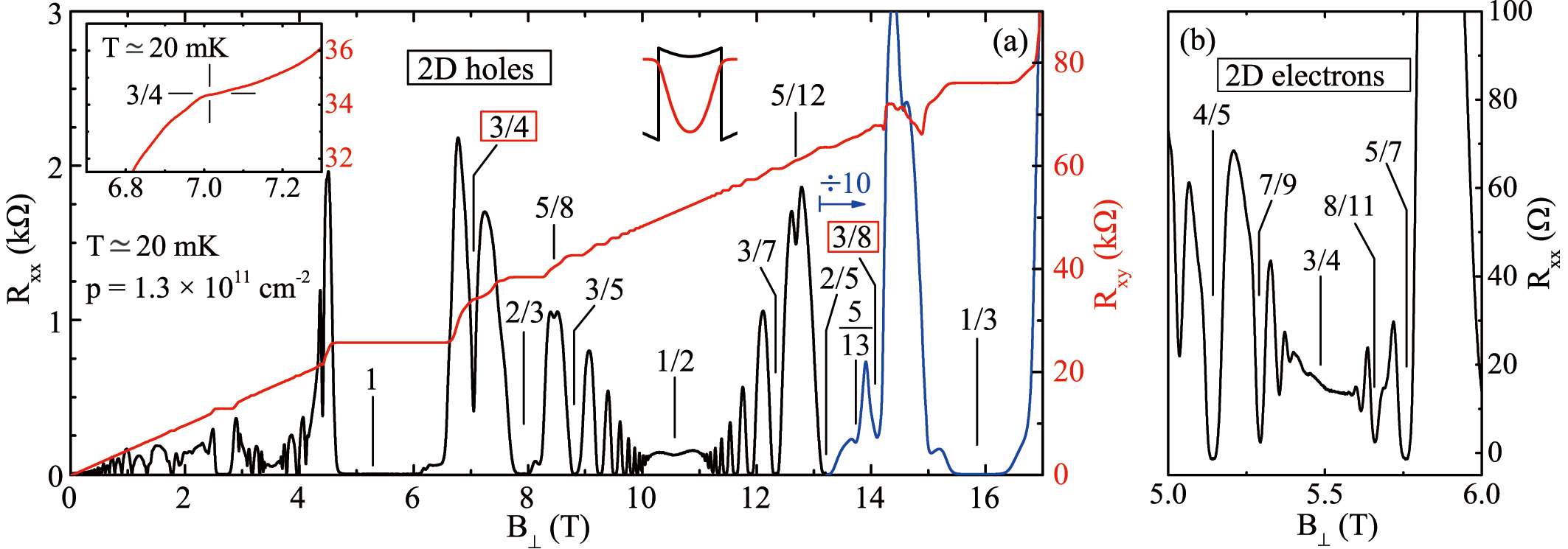, width=1\textwidth}
  \end{center}
  \caption{\label{Fullfield}
    (a) Longitudinal resistance ($R_{xx}$, in black and blue) and Hall resistance ($R_{xy}$, in red) vs perpendicular magnetic field $B_{\perp}$ traces for our ultrahigh-mobility 2D hole sample. The height of the blue trace is divided by a factor of 10. The $B_{\perp}$ positions of several LL fillings are marked. A strong minimum in $R_{xx}$ accompanied by a developing Hall plateau is observed at $\nu=3/4$. An enlarged version of the $R_{xy}$ vs $B_{\perp}$ near $\nu=3/4$ at $20$ mK is shown in the top-left inset. The self-consistently calculated hole charge distribution (red) and potential (black) of the 2DHS are also shown in a top inset. (b) $R_{xx}$ vs $B_{\perp}$ at $T\simeq30$ mK near $\nu=3/4$ for an ultrahigh-mobility 2D \textit{electron} sample with density $1.0\times 10^{11}$ cm$^{-2}$ from Ref. \cite{Chung.NM.2021}. }
  \label{fig:Fullfield}
\end{figure*}

Here we present the experimental observation of an even-denominator FQHS in the lowest LL, at filling factor $\nu=3/4$, in an ultrahigh-quality GaAs 2D \textit{hole} system (2DHS). As highlighted in Fig. \ref{fig:Fullfield}(a), our magneto-transport measurements show a strong minimum in the longitudinal resistance ($R_{xx}$), concomitant with a developing Hall resistance ($R_{xy}$) plateau centered at $(h/e^2)/(3/4)$ to within $0.2\%$. Our finding is unexpected as there is no analogue of such a FQHS in GaAs 2DESs (Fig. \ref{fig:Fullfield}(b)) \cite{Chung.NM.2021} where the ground state at $\nu=3/4$ is a $^4$CF Fermi sea, flanked by odd-denominator FQHSs at nearby fillings (4/5, …, 5/7) which fit into the Jain sequence: $\nu=1-p/(4p\pm 1)$. In our experiments, we find that the $\nu=3/4$ FQHS is fairly robust when a strong in-plane magnetic field is applied, but it eventually gets weaker, with FQHSs at $\nu=4/5$ and $5/7$ emerging on its flanks. We discuss the possible origins of this novel FQHS based on our experimental data and available theories, and suggest that it is likely a non-Abelian state. We also observe a qualitatively similar, but somewhat weaker, FQHS at the even-denominator filling $\nu=3/8$, with likely same origin as 3/4.

The high-quality 2DHS studied here is confined to a 20-nm-wide GaAs quantum well grown on a GaAs (001) substrate \cite{SM, Chung.PRM.2022, Zhu.SSC.2007, Winkler.Book.2003, Alaverdian.PRB.1995, Kevin.PRL.2021}. The 2DHS has a hole density of $1.3\times 10^{11}$ cm$^{-2}$ and a low temperature ($0.3$ K) record-high mobility of $5.8\times 10^{6}$ cm$^2$/Vs \cite{Chung.PRM.2022}. We performed our experiments on a $4$ mm $\times 4$ mm van der Pauw geometry sample. Ohmic contacts were made by placing In/Zn at the sample’s four corners and side midpoints, and annealing at $450$ \textdegree C for 4 min. The sample was then cooled down in two different dilution refrigerators with base temperatures of $\simeq 20$ mK. We measured $R_{xx}$ and $R_{xy}$ using the conventional lock-in amplifier technique, with a low-frequency ($\simeq 13$ Hz) excitation current of $\simeq 10$ nA.

Figure \ref{fig:Fullfield}(a) shows the full-field traces of $R_{xx}$ and $R_{xy}$ vs perpendicular magnetic field $B_{\perp}$ \cite{footnote.isotropic}. The $B_{\perp}$ positions of several LL fillings are marked. A deep minimum in $R_{xx}$ accompanied by a developing $R_{xy}$ plateau is observed at $\nu=3/4$. Weaker $R_{xx}$ minima are also observed at other even-denominator fillings $\nu=3/8$, as well as $5/8$ and $5/12$. Numerous high-order odd-denominator FQHSs are also seen near $\nu=1/2$, up to $\nu=12/25$ and $13/25$, and near $\nu=3/2$, up to $\nu=16/11$ and $17/11$ (not marked in the figure; see Fig. 4 of \cite{Chung.PRM.2022}). These attest to the exceptionally high quality of the 2DHS. Note that the $\nu=3/4$ FQHS is the only FQHS observed between $\nu=1$ and $2/3$. This is in sharp contrast to what is observed in high-quality 2DESs, namely a smooth and shallow $R_{xx}$ minimum with no quantized $R_{xy}$, and flanked by the standard (Jain-sequence) odd-denominator FQHSs such as $\nu=4/5, 7/9, …$ and $5/7, 8/11, …$ (see Fig. 1(b)) \cite{Chung.NM.2021}. It is also in contrast to previous GaAs 2DHSs where, between $\nu=1$ and $2/3$, only weakly-developed FQHSs were observed at $\nu=4/5$ and $5/7$ \cite{Manfra.APL.2005, Liu.PRB.2015}. We also studied another 2D hole sample with higher density, showing weak $R_{xx}$ minima at both $\nu=3/4$ and $5/7$; see SM \cite{SM} for details.

\begin{figure*}
  \begin{center}
    \psfig{file=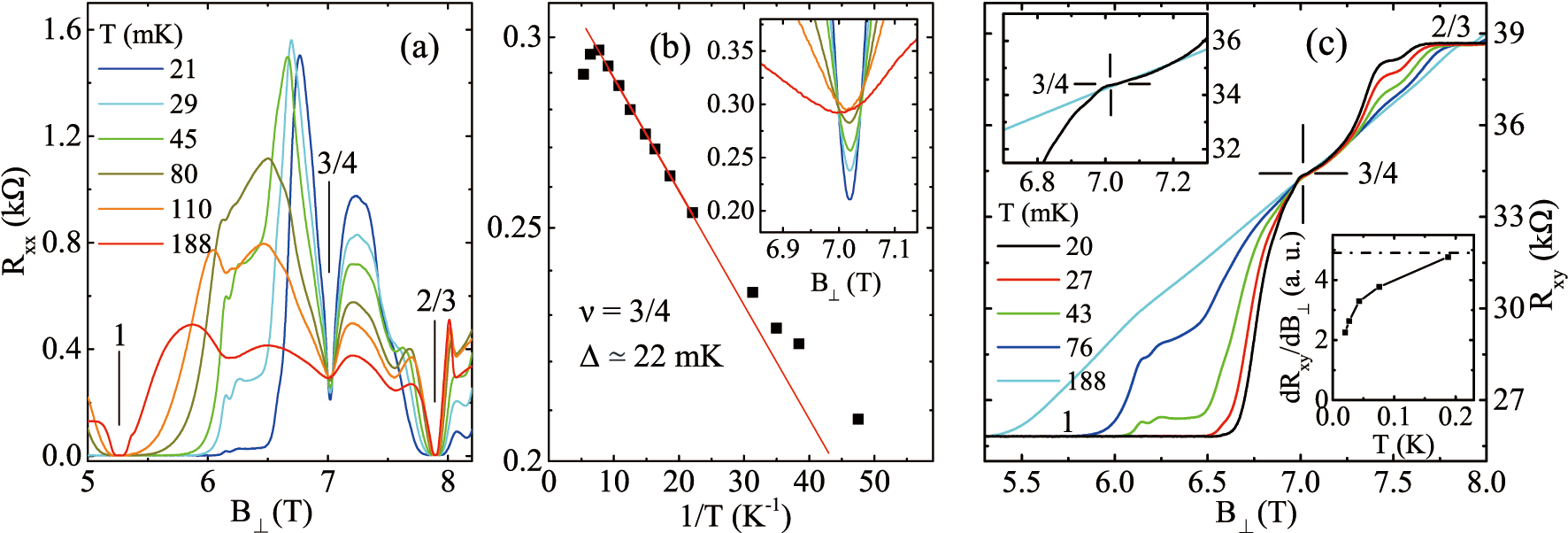, width=1\textwidth}
  \end{center}
  \caption{\label{FQHS} 
     Temperature dependence of $R_{xx}$ and $R_{xy}$. (a) $R_{xx}$ vs $B_{\perp}$ traces near $\nu=3/4$, taken at different temperatures, showing fully developed (nearly zero) $R_{xx}$ minima at $\nu=1$ and $2/3$, as well as a strong minimum at $\nu=3/4$. (b) Arrhenius plot of $R_{xx}$ at $\nu=3/4$. An energy gap of $\simeq 22$ mK is obtained from the linear fit to the data points at intermediate temperatures. Inset: enlarged version of Fig. 2(a) near $\nu=3/4$. (c) Hall ($R_{xy}$) traces taken at different temperatures. $R_{xy}$ is well quantized at its expected value at $\nu=1$ and $2/3$ in the whole temperature range, and shows a developing plateau at $\nu=3/4$ at the lowest temperatures. Top-left inset: enlarged $R_{xy}$ vs $B_{\perp}$ traces near $\nu=3/4$ at $T \simeq 20$ and $188$ mK. Bottom-right inset: Hall resistance slope $dR_{xy}/dB_{\perp}$ vs $T$ at $\nu=3/4$, showing its approach to the expected (classical) value at high $T$ (the dash-dotted line), and to zero as $T$ approaches zero, confirming the $R_{xy}$ quantization.
}
  \label{fig:FQHS}
\end{figure*}

In Fig. \ref{fig:FQHS} we show the temperature ($T$) dependence of $R_{xx}$ and $R_{xy}$ between $\nu=1$ and $\nu=2/3$, measured with a different contact configuration and in a different cool down. When $T$ is reduced, the $R_{xx}$ minimum becomes smaller but its flanks rise steeply, as seen in Fig. \ref{fig:FQHS}(a). In Fig. \ref{fig:FQHS}(b), we show an Arrhenius plot of $R_{xx}$ at $\nu=3/4$. The activated behavior of $R_{xx}$ strongly suggests a FQHS at $\nu=3/4$. An energy gap of $\simeq 22$ mK is deduced from the linear fit to the data points at intermediate temperatures. We note that the temperature range where $R_{xx}$ at $\nu=3/4$ vs $1/T$ follows a linear fit is very narrow. On the low-temperature (large $1/T$) side, the data points start to deviate from the linear fit below $\simeq 30$ mK. Several factors could be causing this deviation: (\textit{i}) At very low $T$, the 2DHS temperature might be slightly higher than $T$ read by the thermometer; (\textit{ii}) $R_{xx}$ at $\nu=3/4$ could be influenced by the rising background on its flanks at very low $T$ (Fig. 2(a)); (\textit{iii}) it is also possible that the deviation is caused by the emergence of a different scattering mechanism (e.g., hopping) at very low temperatures \cite{Boebinger.PRB.1987}. On the high-temperature (small $1/T$) side, the temperature dependence of $R_{xx}$ reverses its trend, and $R_{xx}$ decreases with increasing temperature above $130$ mK. Similar phenomenon was observed for other FQHSs \cite{Samkharadze.PRB.2015}. While the very narrow $T$ range in which we observe an activated behavior in Fig. 2(b) limits the accuracy of the $\simeq 22$ mK energy gap that we determine for the $\nu=3/4$ FQHS, it is likely that this value is an underestimate and is influenced by the strong temperature dependence of $R_{xx}$ on the flanks of $\nu=3/4$. The fact that the $\nu=3/4$ $R_{xx}$ minimum survives at high temperatures (up to $188$ mK) supports this conjecture. 

To further support the presence of a FQHS at $\nu=3/4$, in Fig. \ref{fig:FQHS}(c) we show Hall traces taken at different $T$. $R_{xy}$ is well quantized at the expected values at $\nu=1$ and $2/3$ over the whole temperature range, and shows a nearly quantized plateau at $\nu=3/4$ at the lowest temperature ($\simeq 20$ mK). The top-left inset of Fig. \ref{fig:FQHS}(c) shows an enlarged view of $R_{xy}$ vs $B_{\perp}$ near $\nu=3/4$ at $T \simeq 20$ and $188$ mK. At $\simeq 20$ mK, a plateau occurs at exactly the expected field position of $\nu=3/4$ and is centered at $R_{xy}=(h/e^2)/(3/4)$ to within $0.2\%$. The bottom-right inset of Fig. \ref{fig:FQHS}(c) shows the Hall resistance slope $dR_{xy}/dB_{\perp}$ vs $T$ at $\nu=3/4$. The dash-dotted line represents the expected, classical, high-temperature Hall slope based on the 2DHS density. At low temperatures, $dR_{xy}/dB_{\perp}$ exhibits a trend towards zero, consistent with a developing $R_{xy}$ plateau.

\begin{figure}[t!]
  \begin{center}
    \psfig{file=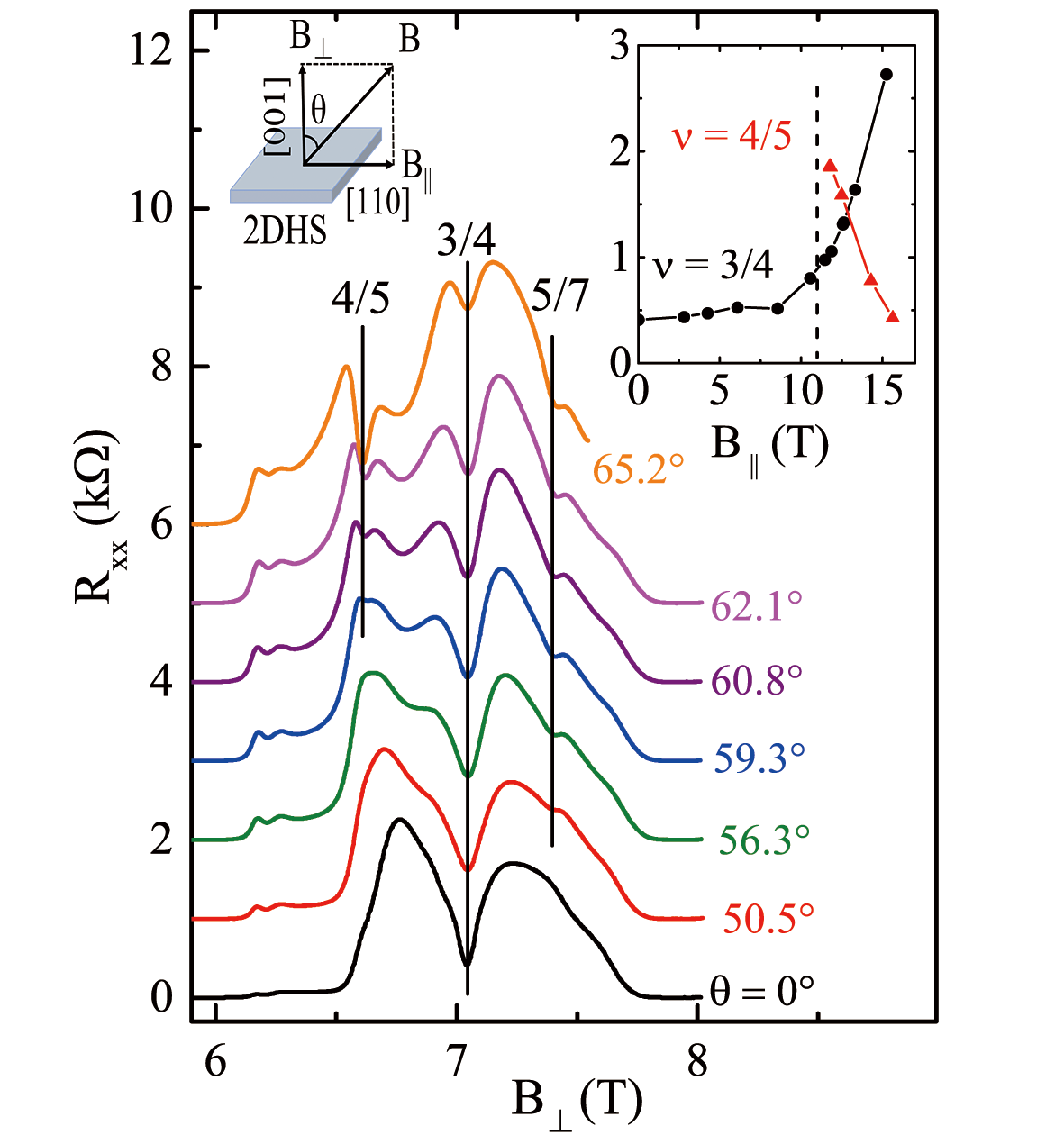, width=0.45\textwidth}
  \end{center}
  \caption{\label{tilt} 
   Tilt angle dependence of $R_{xx}$ is shown vs $B_{\perp}$ at $20$ mK. The left inset shows a schematic of the experimental setup for applying tilted fields; the sample is mounted on a rotating stage to support \textit{in situ} tilt. The traces are vertically shifted for clarity. The vertical lines indicate the $B_{\perp}$ positions for $\nu=3/4$, $4/5$ and $5/7$. The right inset shows $R_{xx}$ vs $B_{\parallel}$ at $\nu=3/4$ (black circles) and $4/5$ (red triangles).
  }
  \label{fig:tilt}
\end{figure}

Next we study the role of an in-plane magnetic field ($B_{\parallel}$) on the $\nu=3/4$ FQHS. Figure \ref{fig:tilt} shows the tilt angle ($\theta$) dependence of $R_{xx}$ vs $B_{\perp}$ between $\nu=1$ and $\nu=2/3$ at $\simeq 20$ mK where $\theta$ denotes the angle between total field ($B$) and its perpendicular component ($B_{\perp}$); see the top-left inset in Fig. 3. The $R_{xx}$ minimum at $\nu=3/4$ remains fairly strong up to $\theta \simeq 60^{\circ}$. With further increase in $\theta$, the $3/4$ FQHS becomes significantly weaker, while a FQHS at $\nu=4/5$ starts to appear and get stronger; see also Fig. \ref{fig:tilt} top-right inset. A hint of a FQHS at $\nu=5/7$ is also observed at large $\theta$, but does not change much with increasing $\theta$. The simultaneous weakening of the FQHS at $3/4$ and appearance of the $4/5$ and $5/7$ FQHSs imply a competition between these states. The origin of this competition is unclear.

Our observation of a $\nu=3/4$ FQHS is unexpected as no such state has been previously seen in experiments or predicted by theory. While a plethora of odd-denominator FQHSs are observed when the Fermi level lies in the lowest ($N=0$) LL, even-denominator FQHSs in single-layer 2DESs confined to different materials such as GaAs \cite{Willett.PRL.1987}, ZnO \cite{Falson.Nat.Phys.2015}, and AlAs  \cite{Hossain.PRL.2018} have been predominantly reported only in the excited ($N=1$) LL. In the $N=0$ LL, in the CF picture and assuming that particle-hole symmetry holds, the ground states at $\nu=3/4$ and $1/4$ are both expected to be Fermi seas of $^4$CFs. In experiments on GaAs 2DESs, a $^4$CF Fermi sea has indeed been directly observed by geometric resonance near $\nu=1/4$ \cite{Hossain.PRB.2019}, and a series of standard (Jain-sequence), odd-denominator FQHSs is seen on the flanks of $\nu=1/4$. A similar sequence of FQHSs is also observed near $\nu=3/4$ in GaAs 2DESs (see Fig. \ref{fig:Fullfield}(b)) \cite{Pan.PRL.2003, Chung.NM.2021}, supporting a $^4$CF Fermi sea ground state. However, in our 2DHS, a CF Fermi sea is not favored at $\nu=3/4$ as evinced by the presence of a FQHS at this filling. The obvious question is: Why is the $3/4$ FQHS observed in our GaAs 2DHS and in the lowest ($N=0$) LL? While we do not have a definitive answer, we discuss below possible explanations.

An unlikely possibility is that the $\nu=3/4$ FQHS in our experiments has an origin similar to the FQHSs reported in the $N=0$ LL of 2DESs in wide GaAs quantum wells at $\nu=1/2$ \cite{Suen.PRL.1992, Suen.PRL.1994, Shabani.PRB.2013, Mueed.PRL.2016} and $\nu=1/4$ \cite{Luhman.PRL.2008, Shabani.PRL.2009, Shabani.PRL.2009.asym}. The origin of these states is in fact still unclear and the possibilities of both a two-component, $\Psi_{331}$, Halperin-Laughlin, Abelian state \cite{Suen.PRL.1994, Shabani.PRL.2009, Shabani.PRB.2013, Peterson.PRB.2010}, and a single-component, Pfaffian, non-Abelian state \cite{Shabani.PRL.2009.asym, Mueed.PRL.2016, Zhu.PRB.2016, Faugno.PRL.2019, Zhao.PRB.2021} have been discussed. Regardless of their origin, these states have only been observed in 2DESs with \textit{bilayer-like} charge distributions. They are also very sensitive to different parameters such as carrier density, quantum well width and symmetry \cite{Suen.PRL.1994, Shabani.PRB.2013}, and magnetic field components \cite{Hasdemir.PRB.2015}. In GaAs 2DHSs, $\nu=1/2$ FQHSs are also seen but, again, only when the charge distribution is bilayer-like \cite{Liu.PRL.2014.1/2}. Our 20-nm-wide GaAs quantum well clearly has a single-layer charge distribution (see Fig. 1(a) inset), and a density well below where a $\nu=1/2$ FQHS is expected based on the phase diagram of Ref. \cite{Liu.PRL.2014.1/2}.  In GaAs 2DHSs, a $\nu=1/2$ FQHS has also been reported when the two lowest-energy LLs cross \cite{Liu.PRB.2014}. As seen in the calculated LL diagram for our sample (see SM \cite{SM}), the crossings between the two lowest-energy LLs are far away from the position of $\nu=3/4$, rendering the crossing origin unlikely. Furthermore, the 3/4 FQHS we observe is quite robust when a strong $B_{||}$ is applied (Fig. 3). In contrast, the FQHS at $\nu=1/2$ in Ref. \cite{Liu.PRB.2014} only appears in a limited range of tilt angles when the crossing occurs close to $\nu=1/2$, and disappears very quickly away from the crossing.

A more likely origin for the emergence of a $\nu=3/4$ FQHS in our 2DHS is the interaction between CFs, caused by the much larger effective mass of holes (compared to electrons) and the ensuing LL mixing. It is well known that CFs are more interacting when the LL mixing is significant, and that the interaction can lead to unusual odd-denominator FQHSs that do not follow the standard Jain sequence \cite{Goldman.Shayegan.Surf.Sci.1990, Pan.PRL.2003, Pan.PRB.2015, Samkharadze.PRB.2015, Chang.PRL.2004, Wojs.PRB.2004, Mukherjee.PRL.2014, Balram.PRB.2021, Balram.PRR.2021}, as well as a Bloch-like spontaneous spin polarization of CFs at low densities \cite{Hossain.Nat.Phys.2021}. Theory in fact predicts that the CF interaction can lead to an even-denominator FQHS at $\nu=3/8$ \cite{Scarola.PRL.2002, Mukherjee.PRL.2012} by mapping this electron filling to the $p=3/2$ effective CF Lambda level ($\Lambda$L) filling of parallel-spin $^2$CFs. The CFs in the half-filled, excited $\Lambda$L capture two additional vortices to turn into $^4$CFs and condense into a paired FQHS where a non-Abelian, anti-Pfaffian state is favored \cite{Mukherjee.PRL.2012}. Note that this $p=3/2$ FQHS would be formed in the excited ($N=1$) $\Lambda$L of CFs, and be equivalent to the celebrated FQHS at $\nu=5/2$ which occurs in the $N=1$ LL of electrons. Experiments have also shown hints of a FQHS at $\nu=3/8$ in GaAs 2DESs, although no conclusive evidence, e.g., a quantized $R_{xy}$, has been reported \cite{Pan.PRL.2003, Bellani.PRB.2010, Chung.NM.2021, Footnote.3/8, Santos.PRB.1992, Ma.PRL.2020}. 

A qualitatively similar scenario can be applied to electrons at $\nu=3/4$ as their filling can be mapped to $p=3/2$ of \textit{anti-parallel} $^2$CFs. (Anti-parallel/parallel here means that the magnetic flux quanta attached to electrons to form $^2$CFs are opposite to/same as the residual magnetic field felt by the $^2$CFs). We note that the $\nu=3/4$ FQHS is spin-polarized, consistent with our observation that the $\nu=3/4$ FQHS is quite robust against $B_{\parallel}$. Similar physics might also explain $R_{xx}$ minima in our data at other even-denominator fillings ($\nu=5/8$ and $5/12$)  as they can also be interpreted by mapping onto $p=5/2$ (anti-parallel and parallel) $^2$CFs, respectively. It is worth noting that in Ref. \cite{Mukherjee.PRL.2012}, the effect of LL mixing was not included, and calculations estimated the gap of the $3/8$ FQHS to be $5$ times smaller than the theoretical gap of the $5/2$ FQHS for a given density. In our 2DHS, however, we see only a hint (weak minimum in $R_{xx}$) of FQHS at $\nu=5/2$, but much stronger $R_{xx}$ minima at $\nu=3/4$ and $3/8$. LL mixing may play an important role in stabilizing the FQHSs at $\nu=3/4$ and $3/8$.

The unexpected $\nu=3/4$ FQHS we observe in a GaAs 2DHS with ultra-high-mobility confirms yet again that the fabrication and availability of samples with unprecedentedly-high quality go hand in hand with the discovery of new interaction phenomena. From this perspective, the GaAs 2D holes, with their large effective mass and unusual LL fan diagram, provide a particularly fruitful platform for exploring novel physics.

\begin{acknowledgments}

We acknowledge support by the National Science Foundation (NSF) (Grants No. DMR 2104771 and No. ECCS 1906253) for measurements, the U.S. Department of Energy Basic Energy (DOE) Sciences Grant No. DEFG02-00-ER45841) for sample characterization, and the NSF (Grant No. MRSEC DMR 2011750), the Eric and Wendy Schmidt Transformative Technology Fund, and the Gordon and Betty Moore Foundation’s EPiQS Initiative (Grant No. GBMF9615 to L.N.P.) for sample fabrication. Our measurements were partly performed at the National High Magnetic Field Laboratory (NHMFL), which is supported by the NSF Cooperative Agreement No. DMR 1644779, by the State of Florida, and by the DOE. This research is funded in part by QuantEmX grant from Institute for Complex Adaptive Matter and the Gordon and Betty Moore Foundation through Grant GBMF9616 to C. W., A. G., and M. S. We thank L. Jiao and T. Murphy at NHMFL for technical assistance, and Jainendra K. Jain for illuminating discussions.

\end{acknowledgments}



\section*{Supplementary material}

# Supplemental Material to "Even-denominator fractional quantum Hall state at filling factor $\nu = 3/4$"

Chengyu Wang,[1] A. Gupta,[1] S. K. Singh,[1] Y. J. Chung,[1] L. N. Pfeiffer,[1]
K. W. West,[1] K. W. Baldwin,[1] R. Winkler,[2] and M. Shayegan[1]

[1]*Department of Electrical and Computer Engineering,
Princeton University, Princeton, New Jersey 08544, USA*

[2]*Department of Physics, Northern Illinois University, DeKalb, Illinois 60115, USA*

(Dated: July 28, 2022)

# I.  SAMPLE PARAMETERS

| Sample | $w$ (nm) | $x_1$ (%) | $x_2$ (%) | $s_1$ (nm) | $s_2$ (nm) | $p$ ($10^{11}$ cm$^{-2}$) | $\mu$ ($10^6$ cm$^2$/Vs) |
|:---:|:---:|:---:|:---:|:---:|:---:|:---:|:---:|
| A | 20 | 16 | 32 | 68 | 100 | 1.3 | 5.8 |
| B | 20 | 26 | 57 | 68 | 100 | 2.2 | 1.8 |

TABLE I. Sample parameters: $w$ is the QW width, $x_1$ is the Al mole fraction of the barrier in the vicinity of the quantum well, $x_2$ is the Al mole fraction of the barrier elsewhere, $s_1$ and $s_2$ are the spacer layer thickness of the barrier with Al mole fraction $x_1$ and $x_2$, respectively, $p$ is the 2D hole density, and $\mu$ is the low-temperature (0.3 K) mobility.

We studied two two-dimensional (2D) hole samples. In both samples, the 2D holes are confined to a 20-nm-wide GaAs quantum well (QW) grown on a GaAs (100) substrate, and are flanked on each side by Al$_x$Ga$_{1-x}$As spacer layers followed by carbon $\delta$-dopings [1]. To achieve high mobility, our samples use a stepped-barrier structure, where the Al fraction of the barrier in the vicinity of the QW is $x_1$ while it is $x_2$ elsewhere [1]. The details of the sample parameters are listed in Table I. Sample A is the one we discuss in the main text. Sample B is a higher-density sample whose data we will present in section III below. The two samples have the same structure except for the Al fractions ($x_1$ and $x_2$) of the stepped barrier. The 2D hole density of sample B is $p = 2.2 \times 10^{11}$ cm$^{-2}$ and the mobility is $\mu = 1.8 \times 10^6$ cm$^2$/Vs at $T \simeq 0.3$ K. Note that the mobility of sample B is significantly lower compared to sample A. Although higher densities are generally beneficial for mobility, several factors could lead to the lower mobility of sample B: ($i$) Based on our band-structure calculations at zero magnetic field, sample B has a larger density-of-states effective mass at the Fermi energy of $\simeq 1.3$ $m_0$ compared to sample A ($\simeq 0.68$ $m_0$), reflecting the highly nonparabolic energy dispersion of these hole systems [2, 3]. (We note that these calculated masses are somewhat larger than the experimentally determined masses. For example, for a 20-nm-wide GaAs QW with a 2D hole density of $1.1 \times 10^{11}$ cm$^{-2}$, cyclotron resonance measurements [2] have indicated an effective mass of 0.54 $m_0$, somewhat smaller than the 0.68 $m_0$ value predicted by our calculations for a similar QW with a density of $1.3 \times 10^{11}$ cm$^{-2}$.) A larger mass leads to a lower mobility as it is inversely proportional to the effective mass. ($ii$) In sample B, the Al fraction in the barrier near the QW is higher, leading to higher impurity concentrations in the barrier which reduces the mobility. ($iii$)

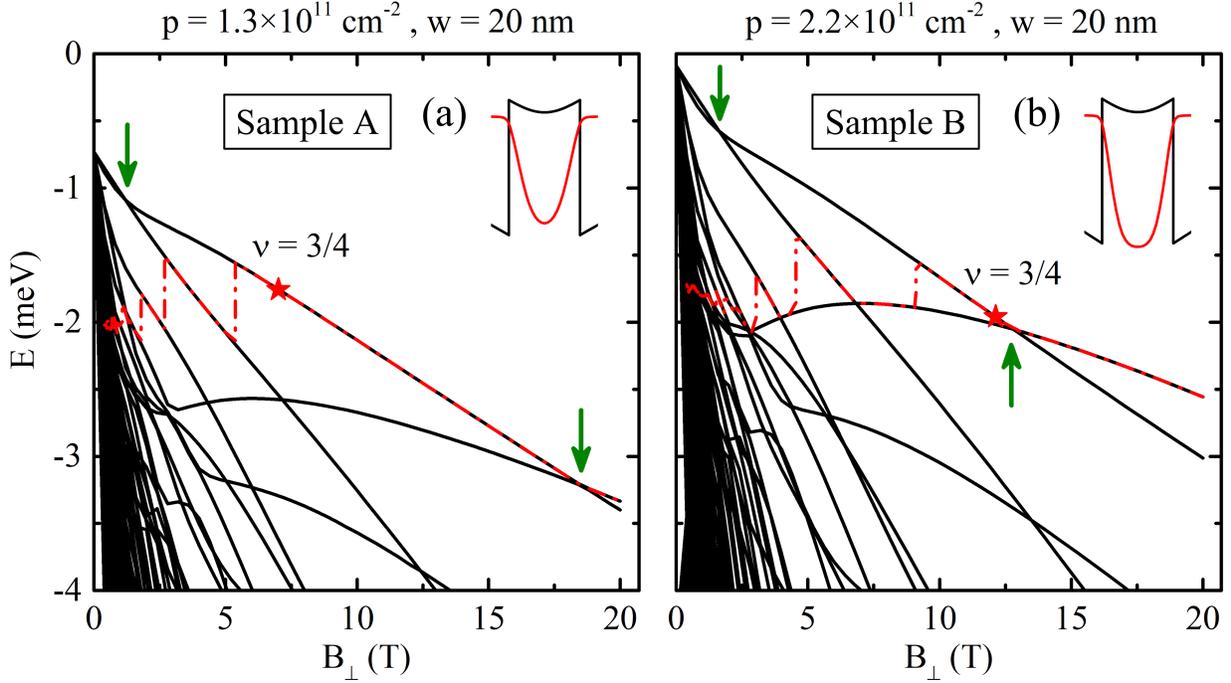

Fig. S1. Calculated energy ($E$) vs perpendicular magnetic field ($B_\perp$) Landau level (LL) diagrams for our 2D hole systems confined to a 20-nm-wide, (001) GaAs QW with a density of (a) $1.3 \times 10^{11}$ cm$^{-2}$ and (b) $2.2 \times 10^{11}$ cm$^{-2}$. Black curves indicate different LLs, and the dash-dotted red curve traces the Fermi energy. In each plot, the field position of $\nu = 3/4$ is marked by a red star, and the green arrows indicate the positions of the crossings of the lowest two LLs. The insets present the potentials (black) and charge distributions (red) of the 2D hole systems.

Sample B likely suffers more from interface roughness scattering because, with a fixed QW width, higher density pushes the hole charge distribution closer to the QW boundaries. The calculated hole charge distributions of both samples are shown in Fig. S1 insets. For a more comprehensive description of optimizing the 2D hole mobility, see Ref. [1].

## II. LANDAU LEVEL CALCULATIONS FOR THE HIGH-MOBILITY TWO-DIMENSIONAL HOLE SYSTEMS

In order to better understand the origin of the fractional quantum Hall state (FQHS) at $\nu = 3/4$ as well as the role of 2D hole density, we calculated the energy ($E$) vs $B_\perp$ Landau level (LL) diagrams for the 2D hole samples we studied. These are shown in Fig. S1. Black

curves indicate different LLs, and the dash-dotted red curve traces the Fermi energy. The calculations were performed using the multiband envelope function approximation based on the $8 \times 8$ Kane Hamiltonian [3]. The insets show that the charge distributions of our 2D hole systems are both single-layer. The two lowest-energy LLs cross twice in the field range we are interested in; see green arrows in Fig. S1. (Here we refer to the LLs with the least negative energy values as having the lowest energy.) The first crossing, occurring at lower field, is between the two lowest LLs with different pseudo-spins. The second crossing, occurring at higher field, is between the two lowest LLs with the same pseudo-spin but opposite parity.

For sample A, the first crossing occurs at $\simeq 2$ T, and the second one at $\simeq 18$ T. These crossings are far away from the field position where the $\nu = 3/4$ FQHS is observed ($\simeq 7$ T). At $\nu = 3/4$, the two lowest-energy LLs are separated by $\simeq 1$ meV. For sample B, the second crossing occurs close to the field position where we see a much weaker $\nu = 3/4$ FQHS, and the two lowest LLs are separated by a much smaller energy ($\simeq 0.1$ meV) at $\nu = 3/4$. We conclude that the crossing of the two lowest LLs is very unlikely to be the origin of the 3/4 FQHS as we observe a particularly strong $\nu = 3/4$ FQHS in sample A.

### III. TRANSPORT DATA FOR A HIGHER DENSITY SAMPLE

Figure S2 presents longitudinal resistance ($R_{xx}$, in black) and Hall resistance ($R_{xy}$, in red) vs $B_\perp$ of sample B. The $B_\perp$ positions of several LL fillings are marked. Between $\nu = 1$ and 2/3, $R_{xx}$ shows a weak minimum at $\nu = 3/4$ accompanied by a hint of developing $R_{xy}$ plateau. A slightly stronger $R_{xx}$ minimum and a developing $R_{xy}$ plateau are seen at $\nu = 5/7$. The inset of Fig. S2 shows the Hall resistance slope $dR_{xy}/dB_\perp$ vs $B_\perp$ at $T \simeq 20$ mK. Minima are seen at $\nu = 3/4$ and 5/7, consistent with developing plateaux at these fillings. Note that the FQHS at $\nu = 3/4$ in this sample is much weaker compared to the lower-density sample in our main text (see Fig. 1), and the 5/7 FQHS observed here is absent in the lower-density sample. We also note that a clearer hint of a FQHS at $\nu = 5/2$ is seen in the higer-density sample.

The reason why the $\nu = 3/4$ and 5/7 FQHSs behave differently in the two samples is unclear. On the one hand, higher 2D hole density pushes the FQHSs at given filling factors to higher fields, which increases the Coulomb energy ($\propto \sqrt{B}$) and thus should strengthen the FQHSs. On the other hand, sample B here has a much lower mobility ($\mu = 1.8 \times 10^6$

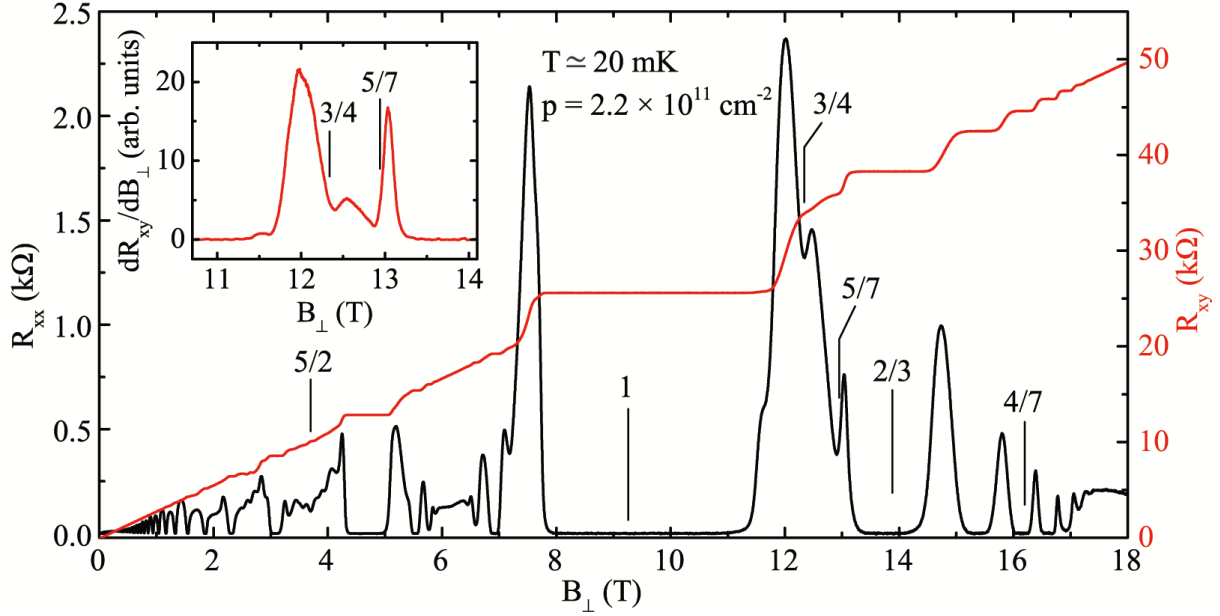

Fig. S2. Longitudinal resistance ($R_{xx}$, in black) and Hall resistance ($R_{xy}$, in red) vs $B_\perp$ of a different 2D hole system, sample B, with higher density. The 2D hole density is $p = 2.2 \times 10^{11}$ cm$^{-2}$ and the mobility is $\mu = 1.8 \times 10^6$ cm$^2$/Vs at $T \simeq 20$ mK. The magnetic field positions of several LL fillings are marked. Inset: Hall resistance slope $dR_{xy}/dB_\perp$ vs $B_\perp$ near $\nu = 3/4$. Positions of $\nu = 3/4$ and $5/7$ in $B_\perp$ are marked.

cm$^2$/Vs) compared to sample A whose data we present in the main text ($\mu = 5.8 \times 10^6$ cm$^2$/Vs), indicating a more disordered 2DHS. It is well known that higher disorder weakens the FQHSs, and this is also consistent with fewer number of FQHSs observed near $\nu = 1/2$ and $3/2$ in sample B despite its higher density.

In addition, LL mixing may also influence the FQHSs. This is usually quantified by the LL mixing parameter $\kappa$, defined as the ratio between the Coulomb energy and the cyclotron energy: $\kappa = (e^2/4\pi\epsilon_0\epsilon l_B)/(\hbar eB/m^*)$, where $l_B = \sqrt{\hbar/eB}$ is the magnetic length and $m^*$ is the effective mass. However, the very complex LL structure and the density-dependent hole effective mass make an assessment of the LL mixing and its consequences very challenging [4]. Previous studies [4] have suggested that a reasonable estimate of $\kappa$ can be made using the 2D hole effective mass at zero magnetic field. Based on the calculated hole effective masses of $\simeq 0.68$ and $\simeq 1.3$ for samples A and B, we estimate $\kappa \simeq 10$ and $14$ at $\nu = 3/4$ for our samples, respectively. Now, the role of LL mixing for the stability of the $\nu = 3/4$ FQHS

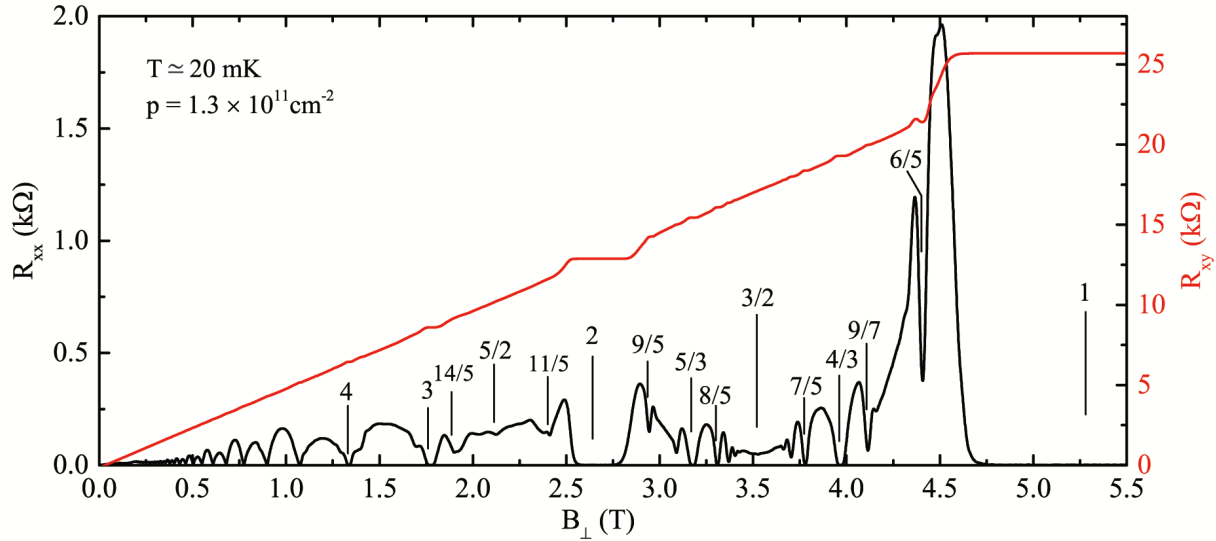

Fig. S3. $R_{xx}$ (in black) and $R_{xy}$ (in red) vs $B_\perp$ of sample A at $\nu \gtrsim 1$. The magnetic field positions of several LL fillings are marked.

is unclear. On the one hand, LL mixing generally weakens the FQHSs [5, 6]. On the other hand, LL mixing also induces residual interaction between composite fermions which could be essential to the emergence of unusual FQHSs, such as one at $\nu = 3/4$.

## IV.  TRANSPORT DATA FOR SAMPLE A AT $\nu \gtrsim 1$

Figure S3 shows the $R_{xx}$ and $R_{xy}$ vs $B_\perp$ of sample A at $\nu \gtrsim 1$. Many odd-denominator FQHSs are seen on the flanks of $\nu = 3/2$ at $\nu = 4/3, 5/3, 7/5, 8/5...$, which belong to the standard Jain sequence states of 2-flux composite fermions. We also observe several FQHSs between $\nu = 1$ and $4/3$, such as at $\nu = 6/5$ and $9/7$. Similar FQHSs are seen between $\nu = 5/3$ and 2, at $\nu = 9/5$ and $12/7$ (not marked in the figure). Between $\nu = 2$ and 3, weak $R_{xx}$ minima are seen at $\nu = 5/2, 11/5$ and $14/5$, however, there is no clear $R_{xy}$ feature at these fillings.

---




\begin{thebibliography}{99}
\bibitem{footnote.isotropic} In our sample, $R_{xx}$ is isotropic in the sample plane.
\end{thebibliography}
\end{document}